# SPACE EFFICIENT CRYPTOGRAPHIC PROTOCOL USING RECURSIVE BITWISE & PAIRS OF BITS OF OPERATION (RBPBO)

**P. K. Jha [1], J. K. Mandal [2]**

[1] School of Engineering and Technology Purbanchal University, Biratnagar, Nepal
E-mail: amp_jha@yahoo.com
[2] Dean, Faculty of Engineering & Management, Kalyani University, Kalyani, West Bengal
Email: jkmandal@rediffmail.com

## Abstract

The technique considers a message as binary string on which a "Efficient Cryptographic Protocol using Recursive Bitwise & pairs of Bits of operation (RBPBO)" is performed. A block of n bits is taken as an input stream, where n varies from 4 to 256, from a continuous stream of bits and the technique operates on it to generate the intermediate encrypted stream. This technique directly involves all the bits of blocks in a boolean operation and a session key. The same operation is performed repeatedly for different block sizes as per the specification of a session key to generate the final encrypted stream.

It is a kind of block cipher and symmetric in nature hence, decoding is done following the same procedure.

A comparison of the proposed technique with existing and industrially accepted RSA and TDES has also been done in terms of frequency distribution and non homogeneity of source and encrypted files.

**Key words:** Space Efficient Cryptographic Protocol using Recursive Bitwise & pairs of Bits of operation (RBPBO), Cipher text, Block cipher, Session Key.

## I. Introduction

In the emerging area of the cryptography [1,2,3,4] strong protocols are used effectively in the strategy of protecting confidential information during its transmission over a network. Information is encrypted at the senders end using an encryption protocol and a key. On reaching at the destination point, the task of decryption is executed using a decryption protocol along with a key to regenerate the source information. Encryption and decryption are in nut shell termed as cryptography. On the basis of the keys used in the entire process, there exists two category of cryptography. In private key cryptography, a single key is used during encryption and decryption and is to be kept private. In public key cryptography, different keys are used during encryption and a private key during decryption.

Section 2 of the paper deals with the principle of this paper. A proposal for key generation and vulnerability is described in section 3. Results are given in section 4. Analysis about the technique is made in section 5. Conclusions are drawn in section 6 and references are drawn in section 7.

## 2. The Technique

This technique operates in two phases:

### a. First phase encrypt the message using Bitwise Operation on Blocks

A stream of bits is considered as the plaintext. the plaintext is divided into a finite number of blocks, each having a finite fixed number of bits like 8/ 16/ 32/ 54/ 255. This technique is then applied for each of the blocks in the following way.

Let $P = s^0_0 \ s^0_1 \ s^0_2 \ s^0_3 \ s^0_4 \ \ldots \ s^0_{n-1}$ is a block of size n in the plaintext. Then the first intermediate block $I_1 = s^1_0 \ s^1_1 \ s^1_2 \ s^1_3 \ s^1_4 \ \ldots \ s^1_{n-1}$ can be generated from P in the following way:

$$s^1_0 = s^0_0$$

$$s^1_3 = s^0_3$$

$s^1_i = s^0_{i-1} \oplus s^0_i$, $1 < i < (n-1)$; $\oplus$ stands for the exclusive-OR operation.

In the same way, the second intermediate block $I_2 = s^2_0 \ s^2_1 \ s^2_2 \ s^2_3 \ s^2_4 \ \ldots \ s^2_{n-1}$ of the same size (n) can be generated by:

$$s^2_0 = s^1_0$$

$$s^2_3 = s^1_3$$

$s^2_i = s^1_{i-1} \oplus s^1_i$, $1 < i < (n-1)$; $\oplus$ stands for the exclusive-OR operation.

If this process continues for a finite number of iterations, the source block P is regenerated forming a cycle, which depends on the value of block size n. Any intermediate block in the recursive process may term as intermediate encrypted block for that source block. The operation is repeated for the whole stream in the source.

**a. Second phase encrypt the output of the first phase using Cascaded Arithmetic Operation on Pairs of bits of a streams.**

This technique, considers the encrypted message from the first phase (here third encrypted block is taken) as a stream of finite number of bits N, and is divided into a finite number of blocks, each also containing a finite number of bits n, where, $1 <= n <= N$.

The rules to be followed for generating cycles are as follows:

1. Consider any source stream of a finite number, where $N=2^n$, $n = 3$ to 8.
2. Make the source stream into paired form so that a pair can be used for the operation.
3. Make the modulo-2 addition (X-OR) between the first and second pair, second and third pair, third and fourth pair and so on of the source stream, to get the first intermediate block.

Any intermediate block in the recursive process may term as intermediate encrypted block for that source block and any block can be taken as the input for the second phase.

Figure 2.1 and fig 2.2 represents the first and second phase of the technique pictorially (single step of the iteration process)

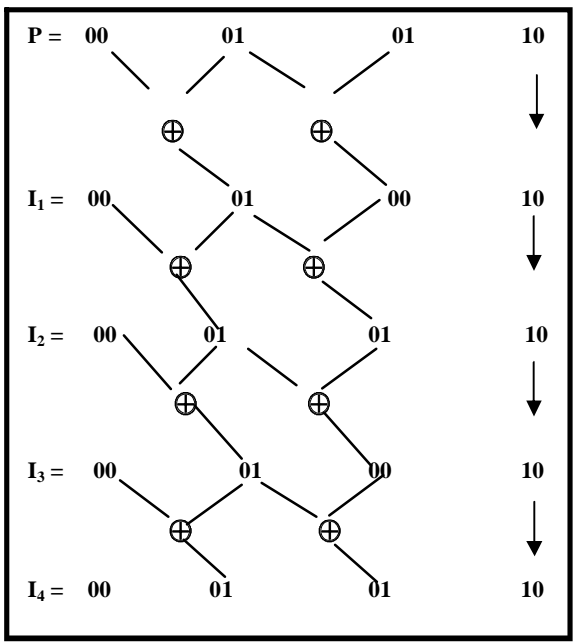

**Figure 2.2: First phase of the technique**

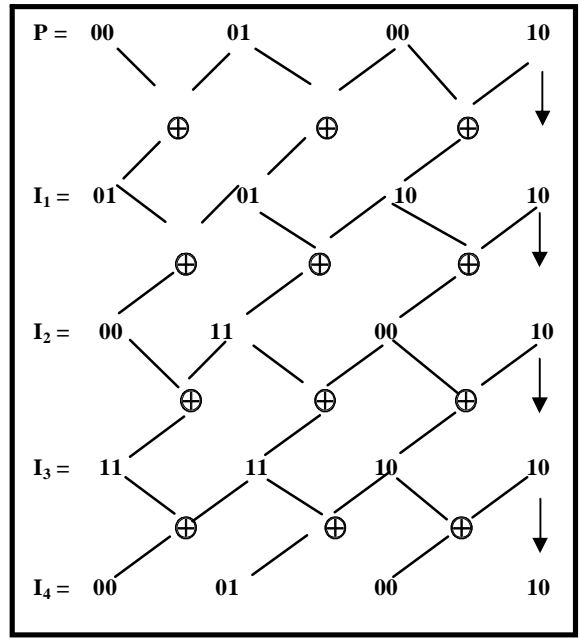

**Figure 2.2: Second phase of the technique**

## 3. Generation of Session Key

To ensure the successful encryption of the proposed technique with varying size of blocks, a 114-bit key format consisting of 12 different segment has been proposed here [1,2,3,4,7,8,9]

For the segment of rank the R, there can exist a maximum of $N=2^{15-R}$ blocks, each of unique size of $S=2^{15-R}$, R starting from 1 and moving till 12.

● Segment with R=1 formed with the first maximum 16384 blocks, each of size16384 bits
● Segment with R=2 formed with the next maximum 8192 blocks, each of size 8192 bits
● Segment with R=3 formed with the next maximum 4096 blocks, each of size 4096 bits
● Segment with R=4 formed with the next maximum 2048 blocks, each of size 2048 bits
● Segment with R=5 formed with the next maximum 1024 blocks, each of size 1024 bits
● Segment with R=6 formed with the next maximum 512 blocks, each of size 512 bits
● Segment with R=7 formed with the next maximum 256 blocks, each of size 256 bits
● Segment with R=8 formed with the next maximum 128 blocks, each of size 128 bits
● Segment with R=9 formed with the first maximum 64 blocks, each of size 64 bits
● Segment with R=10 formed with the next maximum 32 blocks, each of size 32 bits
● Segment with R=11 formed with the next maximum 16 blocks, each of size 16 bits

● Segment with R=12 formed with the next maximum 4 blocks, each of size 4 bits

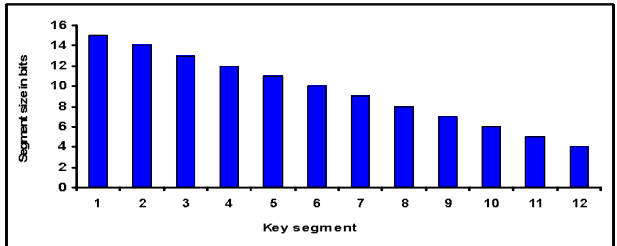

**Figure 3.1**
**114-bit key format of (RBPBO) technique**

With such a structure, the key space becomes 114 bits long and a file of maximum size 42.79 MB can be encrypted using this proposed technique. The key structure is represented in figure 3.1

## 3.1 Vulnerability

Time required to use a brute force approach, which simply involves trying every possible key until an intelligent translation of the cipher text into plain text is obtained. On average, half of all possible key must be tried to achieve success. Table 3.1 shows how much time is involved for various key spaces. Results are shown for four binary sizes. The 56-bit key sizes used with the DES (Data Encryption Standard) algorithm, and the 168-bit key size is used for triple DES. The minimum key size specified for AES (Advanced Encryption Standard) is 128-bits. In the proposed technique key size is 114 bits though this key may be changed to other size as per the requirements.[8,9,10,11]

**Table 3.1**

| Key Size (bits) | No. of alternate Keys | Time required at 1 encryption / µs | Time required at $10^6$ encryption / µs |
|---|---|---|---|
| 32 | $2^{32}=4.3*10^9$ | $2^{31}=35.8$ minutes | 2.15 milli-seconds |
| 56 | $2^{56}=7.2*10^{16}$ | $2^{55}$ µs=1142 years | 10.01 hours |
| 128 | $2^{128}=3.4*10^{38}$ | $2^{127}$ µs=5.4*$10^{24}$ years | 5.4*$10^{18}$ years |
| **114** | $2^{114}=2.07*10^{34}$ | $2^{113}$ µs =3.33*$10^{24}$years | 3.33*$10^{18}$years |
| 168 | $2^{168}=3.7*10^{50}$ | $2^{167}$ µs=5.9*$10^{36}$ years | 5.1*$10^{30}$ years |

**Average time required for exhaustive key search**

## 4. Results

In this section the results of implementations are presented. The implementation is made through high-level language. Table 4.1 represents the encryption time, decryption time, and size before and after encoding and decoding for .CPP files. The encryption time varies from 0.054945 to 0.219780.The decryption time varies from 0.000000 to 0.164835 for the present implementation.

Figure 4.1gives a relationship between encryption times against decryption time. The frequency distribution graphs for source and encrypted files for RBPBO, RSA and TDES techniques are given in figure 4.2.

Chi-square test has also been done and presented in table 4.2 for source files and encrypted files. Results of the Chi square tests are compared with RSA and TDES technique for source files and encrypted files.

**Table 4.1**
**File size V/S encryption/decryption times for .CPP files**

| Source file | Source file size (in bytes) | Encryption time (in seconds) | Output file size (in bytes) | Decryption time (in seconds) |
|---|---|---|---|---|
| VIEWPREV.CPP | 30848 | 0.054945 | 30848 | 0.000000 |
| OLECLI2.CPP | 41023 | 0.054945 | 41023 | 0.054945 |
| OLECLI1.CPP | 61600 | 0.109890 | 61600 | 0.054945 |
| INET.CPP | 72980 | 0.109890 | 72980 | 0.054945 |
| OCCSITE.CPP | 89786 | 0.109890 | 89786 | 0.054945 |
| DBRFX.CPP | 91269 | 0.169890 | 91269 | 0.109890 |
| WINCORE.CPP | 109141 | 0.169890 | 109141 | 0.109890 |
| DBCORE.CPP | 114208 | 0.169890 | 114208 | 0.109890 |
| DAOCORE.CPP | 134431 | 0.219780 | 134431 | 0.164835 |
| BOOK.CPP | 143336 | 0.219780 | 143336 | 0.164835 |

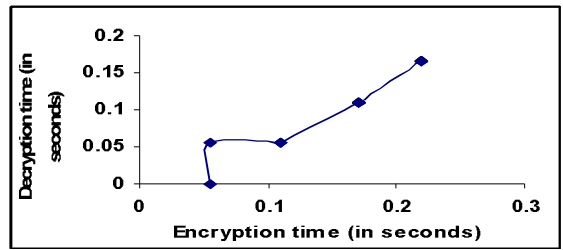

**Figure 4.1**

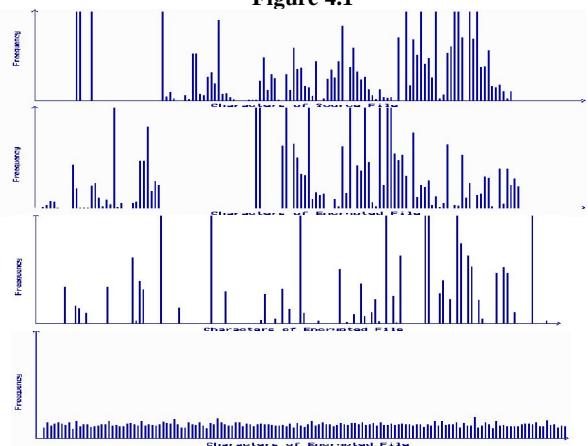

**Figure 4.2**

Graph in figure 4.2 shows the frequency of characters in a message and that frequency of characters in the encrypted message for RBPBO, RSA and TDES technique. The close observation reveals that the character frequencies are more evenly distributed for the RBPBO technique. In connection with the Brute-force attack of decrypting the message by the hackers, it may be difficult to decrypt the message if the message is encrypted using the proposed key system or like manner.

## 4.1 Tests for Homogeneity

The Chi-square test has also been performed using source file and encrypted files for the proposed technique (RBPBO) and existing RSA and TDES technique. Table 4.2 shows the values of Chi-square for different file sizes, which show that the value of Chi-square is almost increasing as file size is increasing. In case of RBPBO the average value obtained for Chi-square test is 50172585. In case of existing RSA technique the average value of Chi-square test for present implementation is 2429648 for the same source file and in case of existing TDES technique the average value of Chi-square test for present implementation is 150257. In all cases the Chi-square is highly significant at 1 % level of significant. So we may conclude that the source files and encrypted files are non-homogeneous in RBPBO with RSA andTDES techniques. The average Chi-square values for RBPBO are greater than RSA and TDES technique in present implementation. Hence, it may be inferred that the technique can be comparable to RSA and TDES technique. Figure 4.3 shows the results of comparison of Chi square values among proposed RBPBO, and RSA, and TDES technique. The bars of blue, gray and yellow shed represent the Chi square for RBPBO, RSA and TDES techniques respectively. It is clear from the table and graph the Chi square values for RBPBO shows either better or comparable result.

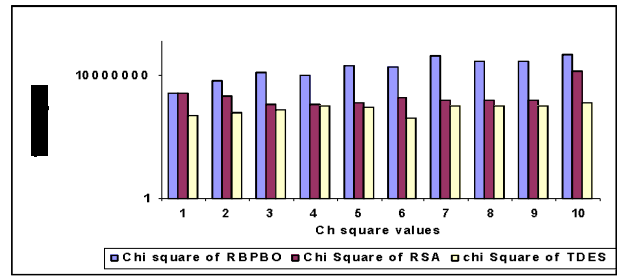

**Figure 4.3**

## 5. Analysis

Analyzing all the results presented in section 4, following are the points obtained on the proposed technique:

- The encryption time and the decryption time vary almost linearly with the size of the source file.
- There exist not much difference between the encryption time and the decryption time for a file, establishing the fact that the computation complexity of each of the two processes is of not much difference.
- For .CPP files there is a relation between the source file size and the Chi square value.
- Chi Square values for .CPP files are very high and vary linearly with the source file size as compared with the other files.
- Out of the different categories of files considered here, Chi Square values for .CPP files are the highest.
- The frequency distribution test applied on the source file and the encrypted file shows that the characters are all well distributed. Chi square values for this proposed technique and those for the RSA/TDES system highly compatible.

## 6. Conclusion

The technique presented here is implemented for different categories of files like .cpp,.exe,.doc,.dll, .sys. When this technique is implemented with X-NOR or other operations using the same logic it will not generate a cycle so this logic cannot be implemented with the other operations. This technique is implemented on 1.3 GHZ processor. In table 4.1 it is seen that as the file size increases the encryption time as well as decryption time increases. For this technique only eight bits blocks are taken, and the third intermediate block is considered here as encrypted stream, so the time required to get the encrypted stream is always be larger than that of decryption because only one iteration is required to get the source stream in the decryption part. Since this technique generates a cycle. It can be easily implemented in any high level language for practical application purpose to provide security in message transmission.

**Table 4.2**

| Source file | Source file in bytes | Value of Chi-square (RBPBO) | Value of Chi-square (RSA) | Value of Chi-square (TDES) |
|---|---|---|---|---|
| VIEWPREV.CPP | 29758 | 1015268 | 1015121 | 54694 |
| OLECLI2.CPP | 39447 | 5207641 | 750711 | 73369 |
| OLECLI1.CPP | 59323 | 16094300 | 215853 | 109858 |
| INET.CPP | 70101 | 9835985 | 223480 | 199820 |
| OCCSITE.CPP | 86262 | 34927048 | 302856 | 159044 |
| DBRFX.CPP | 87810 | 32564780 | 618369 | 35672 |
| WINCORE.CPP | 105194 | 126136000 | 411302 | 194013 |
| DBCORE.CPP | 110526 | 60680092 | 401363 | 204655 |
| DAOCORE.CPP | 129530 | 66645228 | 380307 | 201857 |
| BOOK.CPP | 143336 | 148619507 | 19977116 | 269585 |